\documentstyle[12pt,epsfig,rotate]{article}
\begin{document}

\title
{
EFFECT OF NUCLEAR   TRANSPARENCY  
FROM THE (p,2p) MEASUREMENTS  
ON  $^6$Li AND $^{12}$C AT~1~GeV. \\
}
\author
{  
 { }
 \mbox
 {  V.N.~Baturin, E.N.~Komarov, V.V.~Nelyubin,}\\
 {  V.V.~Sulimov and V.V.~Vikhrov.} \\
\mbox
 {}\\
 {\bf St.-Petersburg Nuclear Physics Institute (PNPI),}\\
 {\bf 188350, Gatchina, Russia.}
}
\maketitle

\begin{abstract}

We studied the  production of protons to the  backward direction   
in $(p,2p)$ reactions on $^6Li$ and $^{12}C$,
accompanied by a  proton emitted into the  forward hemisphere.
The momenta of  the  final two  protons were  measured  in a wide range 
with the  two-arm 
time-of-flight spectrometer. For each event we  reconstructed
the mass of the   intermediate off-shell particles.
We observed that the       
$\Delta(1232)$ is manifested in  the  mass spectra of intermediate baryons.
A surprising  angular dependence of the intermediate meson mass spectrum
was revealed. Also we  have discovered  a strong  narrow  dip   
in the mass spectra of  intermediate mesons at the  mass of the real pion.
We explain this effect  as  an abrupt decrease of the  
absorption probability for  on-shell mesons (the pion-nuclear transparency).

{\it PACS:} 24.30Gd, 25.10+s, 25.75Ld\\

{\it Keywords:} NUCLEAR REACTIONS, 
$^{6}Li$, $^{12}C(p,2p)X$,\\
$P$$=$$1.696GeV/c$, isobar,  
cumulative protons, intermediate particle mass spectrum,
correlation function, nuclear transparency. 
\end{abstract}

\clearpage

\section*{ Introduction. }

One of the 
exciting events in the study  of  
proton-nucleus interactions at the energy scale  of 1GeV
has been the observation of a significant inclusive yield of energetic nucleons
in the energy domain forbidden for  free $pN$ scattering (the cumulative 
effect)\cite{d9}.
Further experiments  revealed 
a  universal exponential behavior of inclusive energy spectra
for various projectiles and outgoing particles   over a wide  energy range
(the so-called nuclear scaling)\cite{leksin}.
Since the momentum transfer can be very high in such processes,
this universality suggests that the cumulative production of nucleons
is relevant to the short-range properties of nuclear matter.
An extensive  theoretical analysis of the    
short-range phenomena 
in hadron- and lepton-nucleus interactions 
in the energy range  from 0.6 to 400GeV has been given in ref.\cite{fs1}. 

\begin{figure}
\begin{center}
\epsfig{file=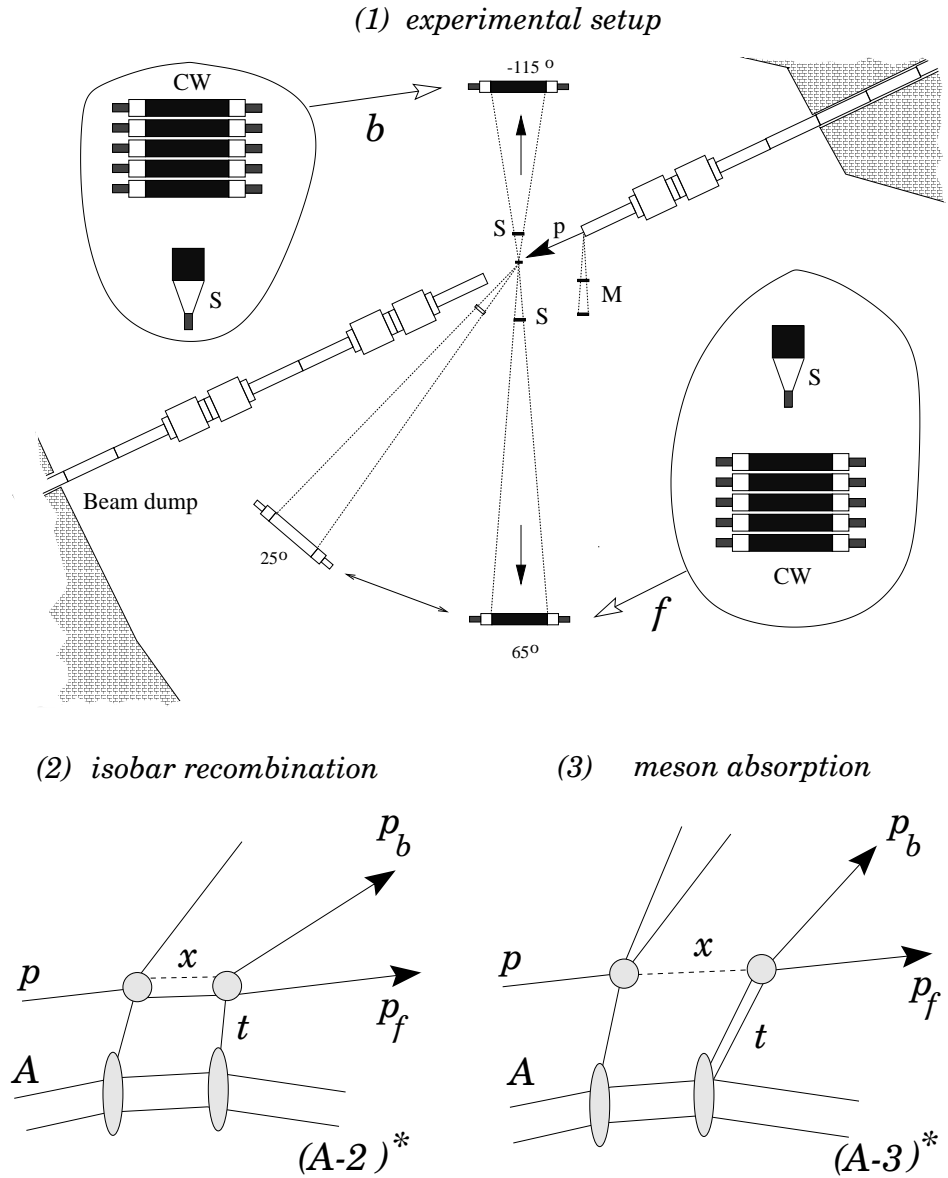,width=13.cm,height=15.5cm}
\end{center}
\caption{ 
(1) Experimental layout: ''p'' - the proton beam,
''M''- monitoring telescope,
''S''-starting counters,
``b''-backward Crystal Wall(CW),
''f''-forward CW.
The inserts show the front view of  CWs and the  starting counters.
(2)The process of isobar ($x$) recombination on a nucleon (t).
(3)The process of meson ($x$)  absorption by a nucleon pair (t).
}
\label{setup}
\end{figure}

In this paper we report on  the emission of  
a backward going  
cumulative proton ($p_b$)
accompanied by a forward  going proton ($p_f$)
\begin{equation}
 p+A \to p_b+p_f+X
\label{reaction1} 
\end{equation}
on $^6Li$ and $^{12}C$ at  a incident proton  energy  1GeV.
Our two arm time-of-flight spectrometer 
allowed the measurement of 
the energies and  the directions of  two  final protons in coincidence.
The experimental layout is shown in Fig.\ref{setup}. The
details of the spectrometer and  the measurements are described in 
Section~1.

From earlier investigations  of  the reaction (Eq.$\ref{reaction1}$)
at 0.64, 0.8 and 1GeV we know that a significant  part of the 
inclusive  spectra  
can be related to phenomena concerning  several  
elementary  collisions,  such as the  mechanism of
quasi-two-body scaling\cite{d38}-\cite{d136},
proton  scattering off a  quasi deuteron pair\cite{d37,d136}, scattering off
 short-range  nucleon pairs\cite{fs1},
and to  the  isobar production in the 
intermediate state\cite{d137}. 
Hence,  the  dynamics  of cumulative production appears complex. 
Nevertheless,  we believe that there is a critical final collision 
in Eq.$\ref{reaction1}$, which results in the  observable protons
\begin{equation}
x+t \to  p_b+p_f, 
\label{reaction2} 
\end{equation}
where $x$  is  an intermediate off-shell particle, and  $t$ is 
a bound target nucleon (or nucleons).
In our study   we focus  on two hypotheses, in which we assume that:
(1)~$t$  is a bound nucleon, while $x$ is an intermediate baryon (isobar) 
and~
(2)~$t$  is a bound nucleon pair, then $x$ has to be an off-shell meson. 
These hypotheses are   referred to  below as ``isobar recombination''
and ``meson absorption''  respectively (see Fig.1).
Of course, the isobar may be produced at the  pion-nucleon vertex.
Therefore, 
we consider the  meson absorption approach as  providing
a further insight  into an  earlier stage of the process,
rather than an alternative to the isobar recombination.

Since  the rest of the struck nucleus has not been detected in our experiment,
the kinematics of the vertex Eq.\ref{reaction2} is under-determined.
Nevertheless,  having measured the  four-momenta of both protons,
 we can evaluate  the invariant mass of  the intermediate particle $x$,
assuming  $t$ to be at rest:
\begin{equation}
m_x^2 = {\textbf{\textit{P}}_x}^2=
(\textbf{\textit{P}}_b +\textbf{\textit{P}}_f - 
\textbf{\textit{P}}_t)^2,
\label{massx01}
\end{equation}
where $\textbf{\textit{P}}_{b,f,t}$ are the four-momenta of final
$b$- and $f$-protons and of  the target nucleon(s) at rest.
Actually{\bf ,}  $t$ is a bound  object in motion, therefore  $m_x^2$ 
has to  be modified by Fermi motion and by  the 
rescattering of the  protons in the final state. 
Using   a  simple  estimate  we show below  that the bias
has to be relatively small for the intermediate baryon. However, it
may be  significant for  the  process of intermediate meson absorption 
by a nucleon pair. 

In the  following Sections~2~and~3
we apply the method of correlation functions 
 to study the  reaction  in Eq.\ref{reaction1}.
First of all, we analyze our data
in terms of  model-independent kinematic variables, such as  the energies
of  secondary particles.
Next we investigate  our first measured mass spectra of intermediate
particles  in both forementioned hypotheses 
and  
discuss the observed effects. 
%

\section{Experimental method.}

The experiment  was performed with  the  two-arm 
time-of-flight spectrometer \cite{d136}-\cite{d143}, which was 
originally designed as a large acceptance neutron detector
for investigation of (p,n) reactions\cite{d139}.
Each arm of the spectrometer consists of a starting scintillator 
(10$\times$10$\times$0.5$cm^3$) 
and a hodoscope of five large scintillators
(20$\times$20$\times$100$cm^3$),
which form the "crystal wall" (CW) with the overall sensitive 
surface of 1$m^2$. Each of the large crystals was viewed  by  
photo-multipliers from both ends.
The starting $b$- and $f$-counters were placed $0.6m$ and $1m$
apart from the target, and they spanned the area of corresponding CWs, placed 
$7m$ and $12m$ away from the target.

The external  proton beam of the PNPI synchro-cyclotron entered the   
target ($10$$\times$$10$$\times$$1cm^3$)  placed in the focus 
of the spectrometer with a yaw minimizing 
the  energy loss of secondary particles.
%
%
The beam 
was monitored by a telescope of two scintillators, 
focused onto the  thin foil window of the vacuum pipe upstream  of
 the target.

\begin{figure}
\begin{center}
\epsfig{file=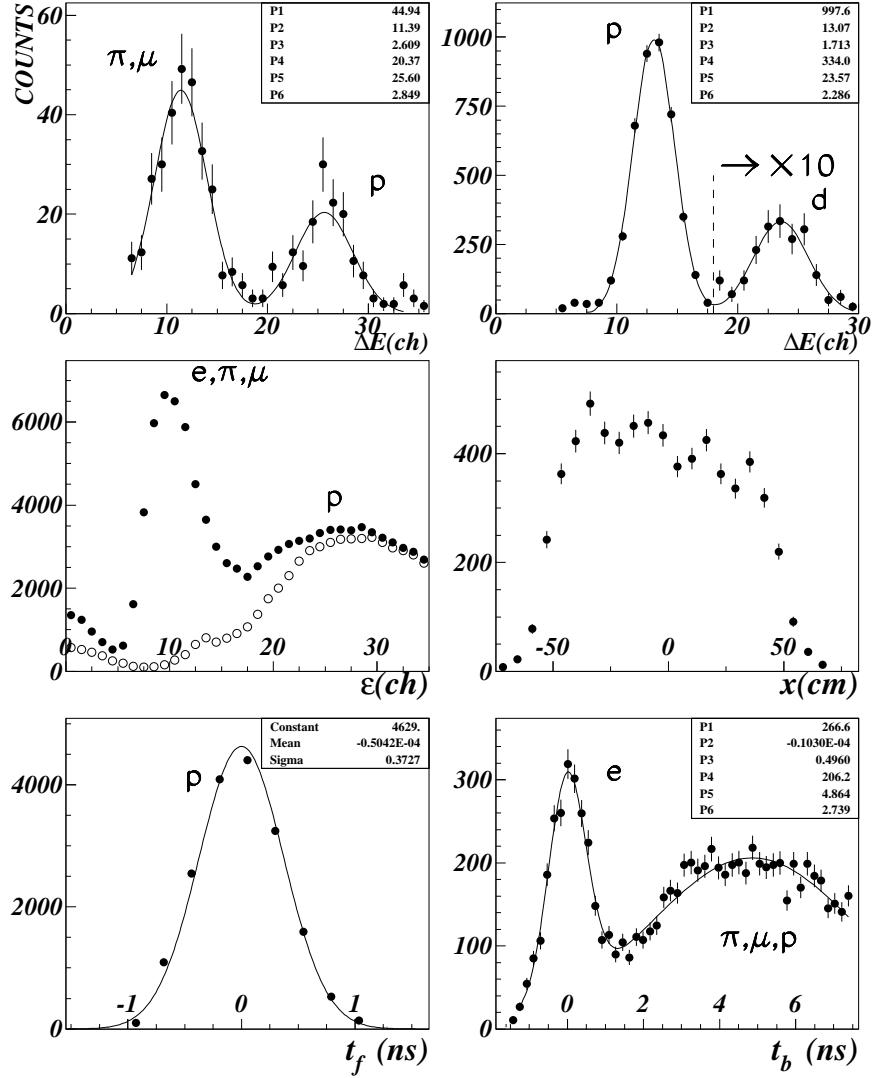,
bbllx=0pt,bblly=128pt,bburx=595pt,bbury=765pt,width=.9\textwidth}
\end{center}
\caption{Distributions relevant to the spectrometer performance. The
panels in this figure are
referred to below as ``row-column''.
(1-1)The sample of energy loss ($\Delta$E) spectrum for 
backward going particles at the velocity corresponding to 
210$\pm10$MeV protons. 
(1-2)The same at the proton energy of 75$\pm$5MeV.
(2-1)Energy loss ($\varepsilon$) in the starting $b$-counter. 
Open circles show the spectrum of particles 
selected at $\Delta$E typical for protons. 
(2-2)Distribution of the $x$-coordinate of backward
going particles.
(3-1)TOF spectrum of 1GeV beam protons. FWHM is $0.87ns$.
(3-2)TOF spectrum of backward going particles at 6.5m
apart the target. FWHM of the electron peak is about $1.2ns$. 
}
\label{tof_bf}
\end{figure}

In order to eliminate the  propagation time of  light in the long scintillators,
the particle  time of flight $t_{i=b,f}$    
has been  evaluated as
$t_{i}=(t_{il}+t_{ir})/2$,
where $t_{il(r)}$ are the time  intervals between the signals from the  
start counters and from the left(right) photo-multipliers of the  scintillator.
Two samples of time-of-flight spectra are shown in  the lower panel of Fig.\ref{tof_bf}.
The resolution in $t_{f}$, resulting from  a set of test beam measurements, 
was  about  $0.87ns$(FWHM). 


The velocities  of secondary protons have been determined via the $t_{b(f)}$
defined above, taking into account  the energy losses along  their 
trajectories.
Since the coordinates of the interaction point in the target are unknown,
the energy loss has been calculated assuming the event originated 
 in the center of the target.
This translates into  an additional smearing of the  particle energy.
The resulting energy resolution in the energy range (100,250)MeV
was between $9\%$ and $4\%$ for $f$-protons  and between $10\%$ and
$8\%$ for $b$-protons. 
At higher energies the resolution for 
$f$-protons  decreased  to $10\%$ at 900MeV.

For the purpose of  momenta reconstruction the 
coordinates of particle hits in both CWs have been  measured.
The vertical coordinate has been determined as the center of the hit 
scintillator in the hodoscope.
The horizontal coordinate $x$ has been measured 
by two  independent methods.
%
Firstly, the  coordinate $x_{t}$ has been measured
via the light propagation time:
2$x_{t}$=$l_0+c(t_{il}-t_{ir})$,
where $l_0$ is the length of the scintillator, $c$ is the speed of
light in the scintillator. 
Secondly, in order to identify  protons by   energy loss,
the signals from the starting counters and from both ends of 
the CWs have been integrated by ADCs. 
The ADC   values  have been  used  
to  measure the  additional coordinate $x_a$  
in a following way.
Due to attenuation of light in the counters, the charges $ q_{l(r)}$   
measured at two ends of the scintillator 
relate to  $x_a$ as 
$q_{l}=k_{l}({\Delta}E){e^{-x_a/l_a}}$ and 
$q_r=k_r({\Delta}E){e^{(x_a-l_0)/l_a}}$, where  
$x_a$ is the coordinate to be measured,
$\Delta$$E$  is the  original
ionization, $k_{l(r)}$  are the  calibration constants,  and $l_a$ is the 
light attenuation length. 
Thus,  both the total energy loss $\Delta$$E$ and  the 
coordinate $x_a$ are known.
A typical   energy loss spectrum of backward going particles
is  shown in the top panel of Fig.\ref{tof_bf}.
The accuracy in the horizontal coordinate combined from the two methods   
$x=(x_a+x_t)/2$,  
measured with direct 1GeV proton beam,
was about  7cm (FWHM).
The resolution in $x$ for secondary 
particles was about 12cm (Fig.\ref{tof_bf}).

The proton beam of the synchro-cyclotron of PNPI  
is delivered on a  sequence of $7ns$ long micro-bunches, 
separated by  $75ns$. 
We have used this peculiarity of  the beam timing
for  a crucial  reduction of  the  background of accidental coincidences. 
The supplementary time interval $t_{bf}$ 
between the hits in starting  $b$- and $f$-counters 
has been digitized for  every event. 
The interval between the birth times of $b$- and $f$-particles
$t^{\ast}_{bf}$ 
has then been  reconstructed\cite{d141},
using the  obvious relation
$t^{\ast}_{bf}$=$t_{bf}$-$\tau_b$+$\tau_f$, where 
$t_{bf}$ 
is the measured time interval, and   
$\tau_{b(f)}$
are the times of particle propagation between  
the target and the  starting counters,
evaluated via $t_{b(f)}$  and the known distances.
Since for real events the final particles  emerge simultaneously, the value  
$t^{\ast}_{bf}$
was found to be distributed  around zero with standard deviation $0.51ns$,
corresponding to the time resolution of the  spectrometer.
By contrast, the  accidental events  filled a  significantly wider  
interval ($14ns$) corresponding to the double width of the micro-bunch. 
We reduced the background of accidental events by an order of magnitude 
by implementing  a cut on $t^{\ast}_{bf}$.
This allowed us to
to  take data at higher beam current
with the same accidental background.

The total  statistics of about $10^5$  true coincidental
events was accumulated during one week run.
While taking data, the $f$-spectrometer has been successively positioned at
$\theta_f$=$12.5^o$, $25^o$$\pm2^o$, $39.5^o$$\pm2.2^o$ 
and $65^o$$\pm2.5^o$, while 
the $b$-spectrometer spanned the area
$\theta_b$ = $-115^o$$\pm4^o$ in the backward hemisphere.

\section {Analysis.}

We have analyzed  our data using the method of  correlation functions (c.f.'s) 
or, in other words, 
the reduced spectra\cite{d136,d137,berger,delphi}.
The substantial advantage of this technique
is that c.f.'s, being the ratios of cross sections, 
are tolerant to  relevant systematics, 
efficiencies, acceptance related effects and diverse   cuts.
The most remarkable  benefit   of this   method 
is  that, as we show below, the deviations
of  reduced spectra may contain a message from the final  vertex.
Thus, provided the  final state and its phase space are unknown, 
this method allows us
to access experimentally the dynamics of the process under study.

In terms of Probability one can
consider the measured differential cross sections 
as a properly normalized probability density functions (p.d.f.'s).
Let  ${\textbf{\textit{b}}}$$\in$$B$ and ${\textbf{\textit{f}}}$$\in$$F$
be the  random multicomponent kinematic variables of 
$b$- and $f$-protons.
Here $B$ and $F$ are the corresponding  manifolds of outcomes.
For inclusive $b$- and $f$-events we 
introduce the 
p.d.fs of  $B$ as
$P_b(\textbf{\textit{b}})$$\propto$$\textit{d}\sigma^i/d\textbf{\textit{b}}$
and of $F$ as 
$P_f(\textbf{\textit{f}})$$\propto$$\textit{d}\sigma^i/d\textbf{\textit{f}}$,
both proportional to the inclusive differential cross sections.
A coincidental event 
(${\textbf{\textit{b}}}$, ${\textbf{\textit{f}}}$)
may be considered as an outcome from the manifold $B$$\cap$$F$ 
with the joint  p.d.f.
$P_{bf}(\textbf{\textit{b}},\textbf{\textit{f}})$
$\propto$
$\textit{d}\sigma^c/d\textbf{\textit{b}}d\textbf{\textit{f}}$,
the double-differential cross section measured in coincidence.
We involve in our analysis  the  combinatoric p.d.f.  
$P^c_{bf}(\textbf{\textit{b}},\textbf{\textit{f}})$=
$P_b(\textbf{\textit{b}})P_f(\textbf{\textit{f}})$,
which has been measured using   
pairs of recorded inclusive $b$- and $f$-events.
We also use the conditional p.d.f.
$Q(\textbf{\textit{b}}|\textbf{\textit{f}})$=
$P_{bf}(\textbf{\textit{b}},\textbf{\textit{f}})/P_f(\textbf{\textit{f}})$,
evaluated at given  $\textbf{\textit{f}}$.
 
First of all, we present our data in terms of the single
particle correlation function $S_f$ given by\footnote
{Similarly, we define $S_b$ with all the following formulas valid 
at $\textbf{\textit{b}}$ $\rightleftharpoons$ $\textbf{\textit{f}}$.}
\begin{equation}
S_f(\textbf{\textit{b}})=
\frac
{Q(\textbf{\textit{b}}|\textbf{\textit{f}})}{P_b(\textbf{\textit{b}})}
\propto
\frac{
d\sigma^c/d\textbf{\textit{b}}d\textbf{\textit{f}}
}
{
d\sigma^i/d\textbf{\textit{b}}
}.
\label{obs113} 
\end{equation}
From the definition of conditional probability one immediately obtains
\begin{equation}
\
\frac{Q(\textbf{\textit{b}}|\textbf{\textit{f}})}{P_b(\textbf{\textit{b}})}=
\frac{P_{bf}(\textbf{\textit{b}},\textbf{\textit{f}})}
{P_b(\textbf{\textit{b}})P_f(\textbf{\textit{f}})}.\
\label{obs112} 
\end{equation}
Two random variables are independent if and only if
$P_{bf}(\textbf{\textit{b}},\textbf{\textit{f}})$=
$P_b(\textbf{\textit{b}})P_f(\textbf{\textit{f}})$=
$P^c_{bf}(\textbf{\textit{b}},\textbf{\textit{f}})$.
Hence, 
due to Eqs. \ref{obs113},\ref{obs112} the c.f.  $S_f$ must be unity,
that is
$Q(\textbf{\textit{b}}|\textbf{\textit{f}})=P_b(\textbf{\textit{b}})$
for  independent emission of $b$- and $f$-protons.
The last relation may be integrated over any number\footnote{
In the following formula we integrate it over all 
but one components of \textit{b}.}
of components of 
$\textbf{\textit{b}}$ and  $\textbf{\textit{f}}$.  
Applied to the measured spectra, this leads 
to an important  statement:
for  independent emission
of particles
\begin{equation}
S_f({\textit{b}})\propto\frac
{(d\sigma^c/d{\textit{b}})_f}
{(d\sigma^i/d{\textit{b}})}
=const(b),
\label{obs13} 
\end{equation}
where $b$ is a  scalar 
variable related to the $b$-proton,
such as  the kinetic energy. The index $f$ means that the cross section
is integrated over $\textbf{\textit{f}}$$\in$${f}$$\subset$${F}$.
Thus, if the  measured c.f.  $S_f(b)$ 
deviates from a constant\footnote{Absolute normalization of $S$ 
is not important.},
then, very likely, 
a signal from a common vertex  of  
$b$- and $f$-protons is being observed.

A particularly promising way of studying the process
Eq.\ref{reaction1}
is by means 
of the mass spectra of intermediate particles.
%
Let us consider the c.f. constructed from the  inclusive
and  double differential cross sections:
%
\begin{equation}
R(\textbf{\textit{v}})=\frac
  {P_{bf}(\textbf{\textit{v}})}
{P^c_{bf}(\textbf{\textit{v}})}
\propto
\frac{d\sigma^c/d\textbf{\textit{b}}d\textbf{\textit{f}}}
{(d\sigma^i/d\textbf{\textit{b}})(d\sigma^i/d\textbf{\textit{f}})},
\label{obs101} 
\end{equation}
where $ P_{bf}$ and $P^c_{bf}$ are  the  coincidental and 
combinatoric p.d.f's introduced above.
We  
consider them to be functions of the compound vector  
$\textbf{\textit{v}}=(\textbf{\textit{b}},\textbf{\textit{f}})$.
The ratio of density functions $R(\textbf{\textit{v}})$ 
can be expressed via a new variable
$\textbf{\textit{u}}=
\textbf{\textit{u}}(\textbf{\textit{v}})$,
using the technique of Jacobians,  which   obviously are canceled in the ratio. 
If the final  protons emerge  independently, then again 
$P_{bf}(\textbf{\textit{u}})=P^c_{bf}(\textbf{\textit{u}})$. 
One may integrate this relation over any
components of  $\textbf{\textit{u}}$, 
and  we obtain\footnote{Integrating over all, 
but one components of \textbf{\textit{u}}.}
another useful statement, valid for independent emission
\begin{equation}
R(u_i)=\frac{P_{bf}(u_i)}{P^c_{bf}(u_i)}=1\propto{const(u_i)},
\label{obs141} 
\end{equation}
where
$P_{bf}(u_i)=\int\Pi_{{j}\neq{i}}du_{j}{P^c_{bf}(\textbf{\textit{u}}})$
and $u_i$ is a scalar variable, such as  invariant mass of an intermediate 
particle.
Thus, a deviation of the measured two particle c.f. $R(m_x)$ 
from a constant\footnote
{We do not normalize the measured $R$  to unity since the
normalization is not important.}
may be considered  as an indicative of an intermediate particle $x$.

\section{Results and discussion.}

\begin{figure}
\begin{center}
\epsfig{file=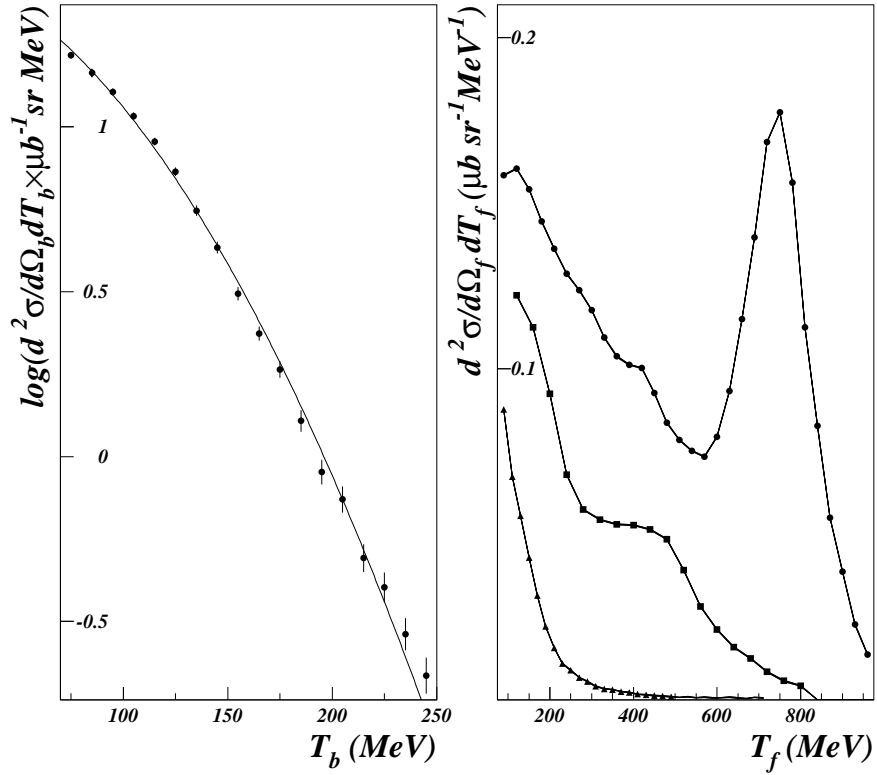,
bbllx=0pt,bblly=128pt,bburx=595pt,bbury=765pt,width=.9\textwidth}
\end{center}
\caption{Kinetic energy spectra of inclusive $b$- and $f$-protons from
the $^6Li$ target.
Left: $T_b$ spectrum at $115^o$.
Right:
(1)circles - $T_f$ spectrum at $\theta_f$=$25^o$. 
Peak at  745MeV(sigma 69MeV) corresponds to quasi-elastic $pN$-scattering. 
(2)squares -  at $39.5^o$. The quasi-elastic peak is evident at  470$\pm$
10MeV (sigma 79MeV).
(3)triangles - at  $65^o$. Solid lines are shown to guide the eye.}
\label{incl_pic}
\end{figure}

\begin{figure}
\begin{center}
\epsfig{file=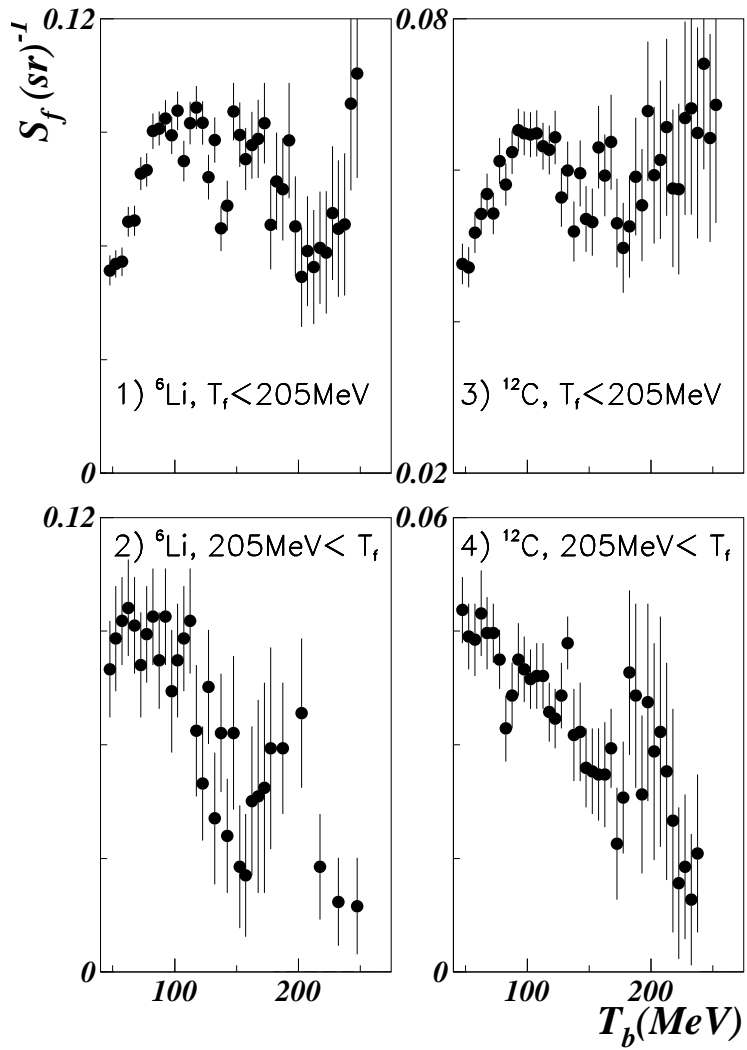,
bbllx=0pt,bblly=128pt,bburx=595pt,bbury=765pt,width=.9\textwidth}
\end{center}
\caption{Correlation function $S_f$ for target $^{6}Li$  at
$\theta_f=65^\circ$ 
vs $T_b$, the  kinetic energy of backward going proton:
(1) at $T_f$$\le$205MeV and (2) at $T_f$$>$205MeV.  
The same for target $^{12}C$: (3) at $T_f$$\le$205MeV and
(4) at $T_f$$>$205MeV.}
\label{dmb_c12+li6}
\end{figure}

The inclusive energy spectra of $b$- and $f$-protons,
used for  evaluation of  c.f.'s,
are shown in Fig.\ref{incl_pic}.
First we focus on c.f. $S_f$, measured
at $\theta_f$$=$$65^o$ and  shown in  Fig.\ref{dmb_c12+li6}
as a  function of kinetic energy of $b$-protons~($T_b$).
In this figure  $S_f(T_b)$ scales in units of  
differential multiplicity, i.e. having multiplied $S_f$ 
by the solid angle of the $f$-spectrometer, one obtains the mean
number of $f$-protons  accompanying the single $b$-proton of energy $T_b$.

\begin{figure}
\begin{center}
\epsfig{file=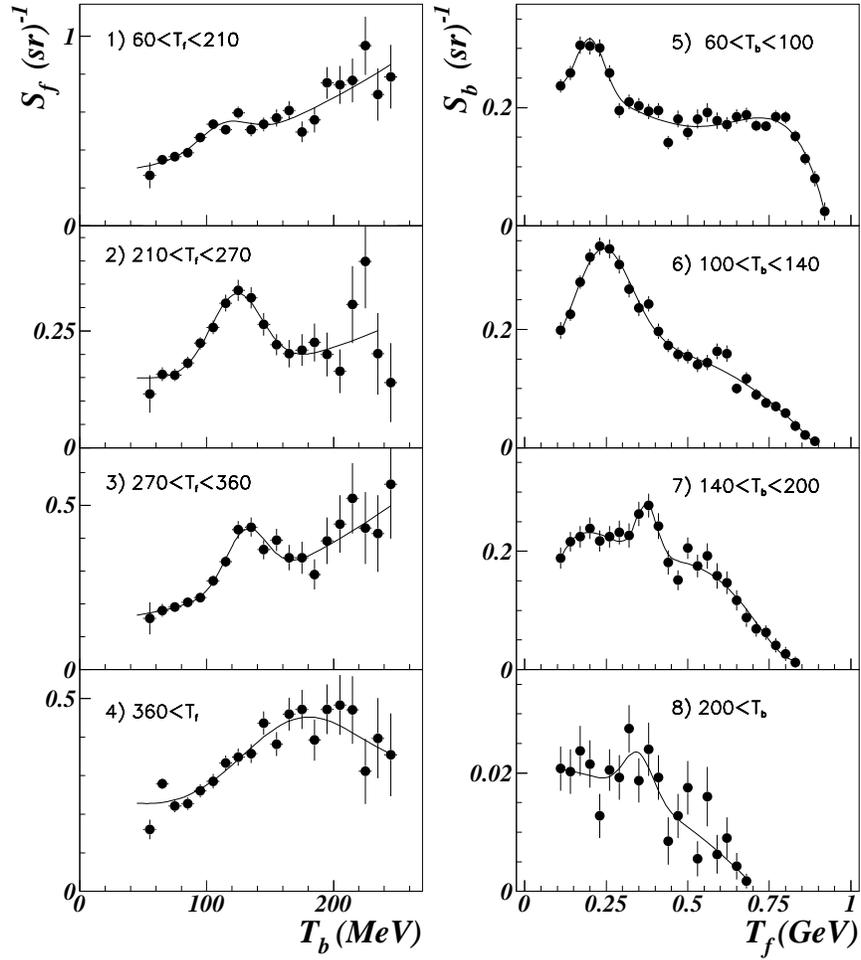,
bbllx=0pt,bblly=128pt,bburx=595pt,bbury=765pt,width=.9\textwidth}
\end{center}
\caption{Cross reference study of  $S_b$ and $S_f$ 
for $^{6}Li$ at $\theta_f$=$25^\circ$.
(1-4)$S_f(T_b)$ 
for different intervals of $T_f$ specified in each fragment.
(5-8)$S_b(T_f)$ for different intervals of $T_b$.
Solid curves  are the  fits by a Gaussian plus polynomials.}
\label{crossreference25g}
\end{figure}

Noticeable deviations   from  a constant are  seen
in the vicinity of 100 MeV on both targets. 
The relative amplitude of the deviation is obviously $A$-dependent. 
The shape of $S_f$   varies with the
energy of the $f$-proton. 
To further investigate  these deviations, the
reduced energy spectra of both $b$- and $f$-protons have been studied 
 for their dependence on
the energy of the partner. 
The example of such a ``cross-reference'' study 
at $\theta_f$=$25^o$
is shown in Fig.\ref{crossreference25g}. 
A narrow peak  shows up at  $115$$\pm$$4$MeV in the shape of $S_f(T_b)$ 
when the energy of  $f$-proton increases\footnote{The numbers are from
a fit by the sum of a Gaussian and polynomials.}. 
The amplitude of this peak reaches a  maximum at $f$-proton energy in
the interval  $(210,270)$MeV.
The deviation is also evident  in  $S_b(T_f)$ as  a bump at  $f$-proton energy
$237\pm10$MeV. 
%
Both fitted ``characteristic'' energies
are in agreement with the   
kinematics of the  two-stage intra-nuclear process
$pp_1\to\Delta^+p$,
$\Delta^+$$p_2$$\to$$p_bp_f$ involving the slow\footnote{
The  ``slow''(``fast'') isobar may be produced
on the target(projectile) nucleons.} isobar 
and the protons ($p_{1,2}$) from the nucleus\cite{d137}.
We note that the sum of characteristic energies ($352$MeV) 
corresponds to a kinetic energy of the isobar of $58$MeV. 
If scaled to the beam energy, this  value agrees with the measured
spectrum of  isobars from 
$pp\to\Delta^+p$ at 1.47GeV\cite{P1747-62}
which has a maximum  corresponding to an isobar energy of 80MeV.

The cross reference
study had been  also  performed  at $\theta_f=39.5^o$ and $65^o$,  
and similar indications of the isobar recombination process 
were observed\cite{d137}.
We find  it  fascinating  that the  isobar recombination mechanism 
is manifest even in the reduced spectra of model independent
variables.

\begin{figure}
\begin{center}
\epsfig{file=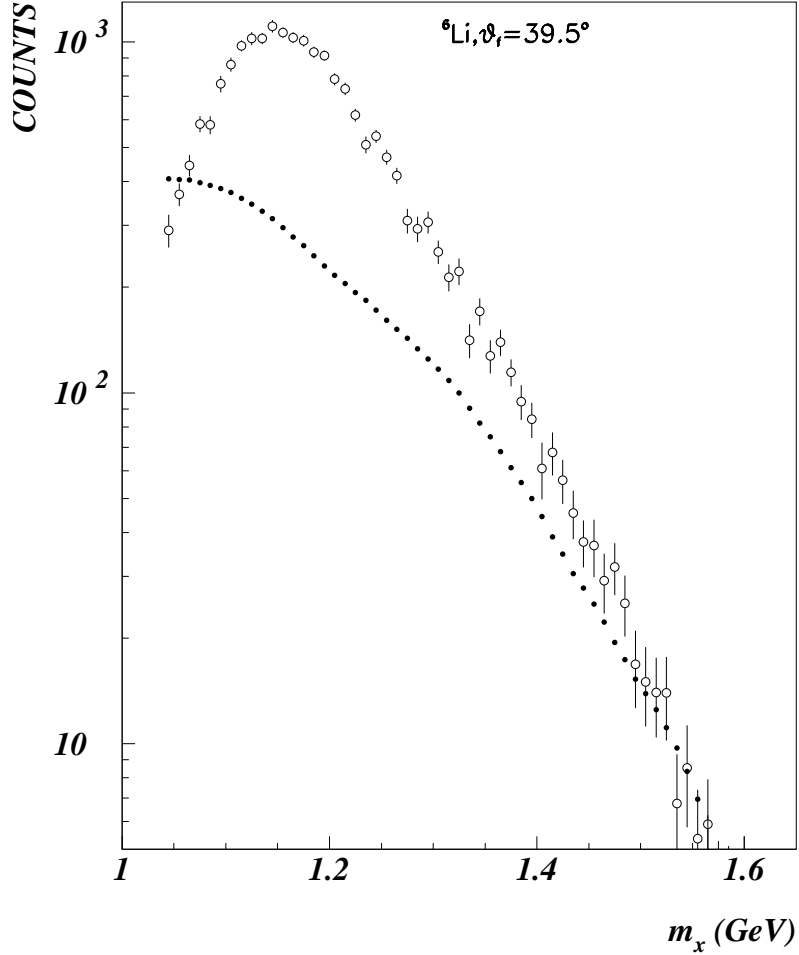,
bbllx=0pt,bblly=128pt,bburx=595pt,bbury=765pt,width=.9\textwidth}
\end{center}
\caption{The spectra of intermediate baryon mass($m_x$) from
$^{6}Li$
at $\theta_f$$=$$39.5^\circ$.  
The mass  is calculated  within  the 
$\Delta$ recombination hypothesis.
The open circles show the raw mass spectrum. The black points
represent the 
combinatoric spectrum with an 
arbitrary normalization. The statistical errors are less than the
symbol size.}
\label{delta_acc}
\end{figure}

\begin{figure}
\begin{center}
\epsfig{file=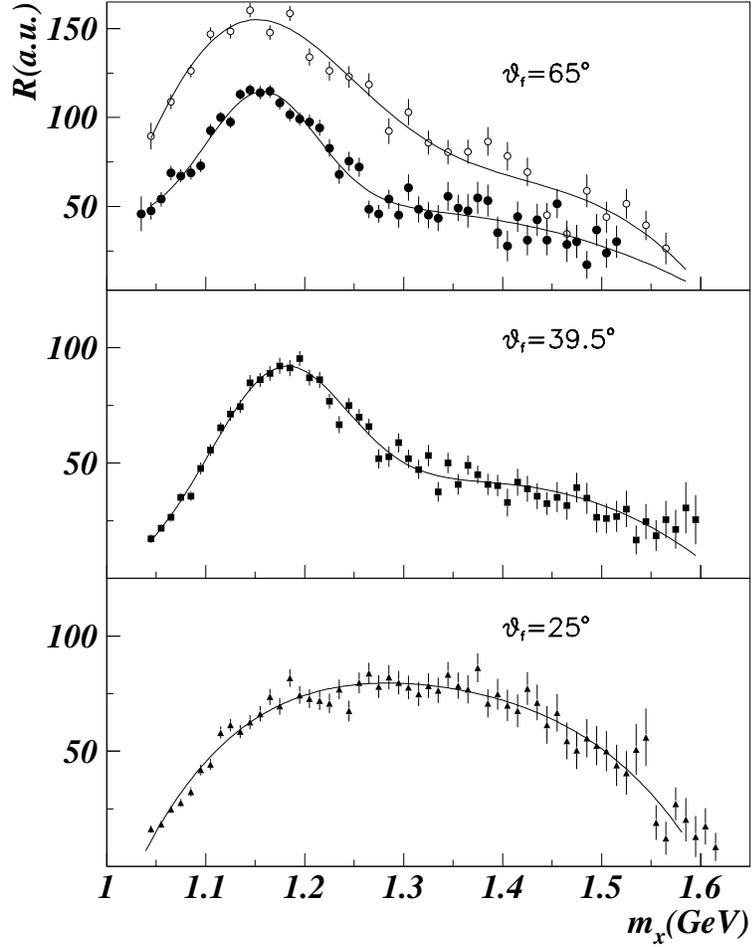,
bbllx=0pt,bblly=128pt,bburx=595pt,bbury=765pt,width=.9\textwidth}
\end{center}
\caption{Correlation function $R$ on targets $^{6}Li$ (filled symbols)
and $^{12}C$(open circles) vs $m_x$, the     
mass of intermediate baryon  ,  calculated within the isobar 
recombination hypothesis at  $\theta_f$ :
(1)$65^\circ$(circles),
(2)$39.5^\circ$(squares),
(3)$25^\circ$(triangles).
Solid curves are the  fits by a Gaussian plus polynomials.}
\label{f_li_deltamass}
\end{figure}

 Other very interesting observations can be made on   
the mass spectra of intermediate particles\cite{d136,d137}.
An example of the initial intermediate baryon  mass spectrum 
is shown in Fig.\ref{delta_acc} in 
comparison with the corresponding combinatoric spectrum.
The two-particle correlation function $R$ v.s. $m_x$ (the  
intermediate baryon mass) is shown in Fig.\ref{f_li_deltamass}
at three angles of the forward-going proton. 

A strong deviation of  $R(m_x)$ is  distinctly seen in  
Fig.\ref{f_li_deltamass} at  $\theta_f$=$39.5^o$ and $65^o$ 
as a peak in the vicinity  of the $\Delta(1232)$ resonance.
The centroids (1173$\pm$3, 1156$\pm$3MeV) and the widths (150$\pm$10, 
130$\pm$10MeV) obtained in a fit are close to  the parameters of the isobar. 
The   isobar peak  is quite prominent  at $39.5^o$ and $65^o$, but  
at  $25^o$ it is hidden by the events of the  quasi two-body 
scattering\cite{d136}, which are spread between $1.2$ and $1.6$GeV.
Nevertheless, some  indications of the isobar at $25^o$ have already been  
shown in Fig.\ref{crossreference25g}.

\begin{figure}
\begin{center}
\epsfig{file=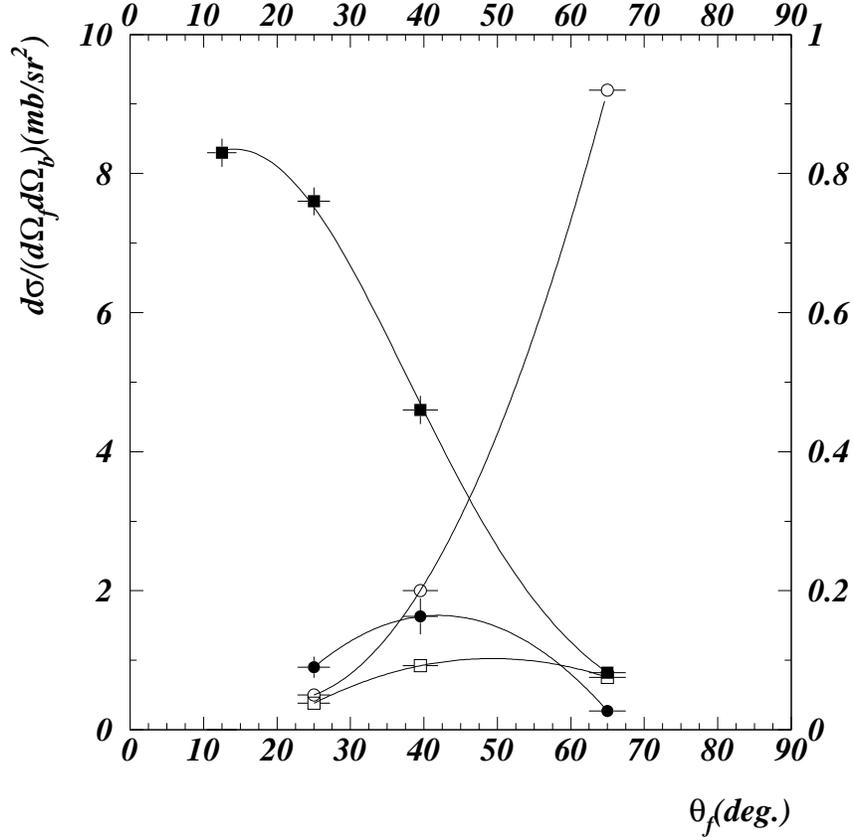,
bbllx=0pt,bblly=128pt,bburx=595pt,bbury=765pt,width=.9\textwidth}
\end{center}
\caption{
Cross sections vs the angle  
of the  forward-going proton($\theta_f$). 
(1)~The total yield  of the coincident $b$- and $f$-protons (filled squares).
(2)~The yield from the $\Delta$ recombination process (filled circles).
(3)~The yield of positive $m_{x}^2$ within the meson absorption hypothesis 
(open squares).
(4)~The ratio of the positive $m_{x}^2$ yield  to 
the total yield (open  circles, right scale).
Shown by solid curves are the fits by polynomials to guide the eye.}
\label{angular_dependence}
\end{figure}

From the behavior of  both  $R$ and $S$ 
we have estimated the $\theta_f$-dependence of the 
$\Delta$ recombination yield, 
which is plotted in Fig.\ref{angular_dependence}.
Integrating this curve over $\theta_f$ and the azimuth\footnote{
We assume the azimuthal distribution to be uniform.}, 
we have roughly estimated the total yield of    
$\Delta$-recombination process 
to be about 20\% of the inclusive yield at $115^o$.

There are several  experimental indications 
that nuclear medium effects do modify the 
parameters of the $\Delta$ resonance\cite{electrodelta}.
With this concern we notice that the centroids of the 
isobar peak at $39.5^o$ and $65^o$ (Fig.\ref{f_li_deltamass})
are displaced to lower mass by $59$$\pm$$3$ and $76$$\pm$$3$MeV respectively.
%

To estimate  the  influence  of  Fermi motion 
on the position of the isobar peak on the  mass scale  
one may  compare the yield of Eq.\ref{massx01} at    
$\textbf{\textit{P}}_{\tiny{t}} = (m_{\tiny{t}} ,\vec{0})$
with  the  value 
$ m{^\prime}_x^2$,
calculated for an off-shell object $t$ in motion, i.e. at 
$\textbf{\textit{P}}_{\tiny{t}}$=
$\textbf{\textit{P}}^{\prime}_{\tiny{t}}$
=$(E_{\tiny{t}} ,\vec{p}_{\tiny{t}})$.
In every event the difference of these values  is given by
\begin{eqnarray}
\nonumber
{
m_x^2 - m{^\prime}_x^2 
= (\textbf{\textit{P}}{^\prime}_t-\textbf{\textit{P}}_t)
(2(\textbf{\textit{P}}_b+\textbf{\textit{P}}_f)
 -(\textbf{\textit{P}}{^\prime}_t+\textbf{\textit{P}}_t))}= \\
{2(E_t-m_t)(E_b+E_f)-2(\vec{p}_t,\vec{p}_b+\vec{p}_f)-E_t^2+
\vec{p}_t^2+m_t^2}.
\label{massx02}
\end{eqnarray}
Considering Eq.\ref{massx02} as  the operator on  vector 
$\mid$$t$$\rangle$,  
one obtains the eigenvalue of 
$\hat{m}{^\prime}_x^2$ 
in the form 
\begin{equation}
\langle\hat{m}{^\prime}_x^2\rangle-m_x^2
=-2\varepsilon_t(E_b+E_f-m_t)-\langle\hat{p}_t^2\rangle+\varepsilon_t^2,
\label{deltamass2}
\end{equation}
where $\varepsilon_t=\langle\hat{E}_t\rangle-m_t$ is the binding energy.
We estimate the value of  $\langle\hat{p}_{t=N}^2\rangle$ via  the
wave functions\footnote{
${\mid}1P_{3/2}{\rangle}$:
$\varepsilon_{N}$=$-5.8$MeV,
$\langle\hat{p_{N}}^2\rangle$=$.028$GeV$^2$.
${\mid}1S_{1/2}{\rangle}$:
$\varepsilon_{N}$=$-23$MeV,
$\langle\hat{p_{N}}^2\rangle$=$.017$GeV$^2$.}
of  $^6Li$ given in ref.\cite{Mikl1}, and we find from   Eq.\ref{deltamass2},
at ``characteristic'' values of   $E_b$ and  $E_f$,
that  Fermi motion on  $P$- and $S$-shells
may shift the mass 
by  
$-5.5\pm0.3$MeV
and 
by 
$17\pm1$MeV, 
respectively\footnote{
The error bars here account  for the width of the isobar.}.
Hence, the contribution from  Fermi motion can not be entirely  
responsible for the observed displacement. 

The mass spectrum of intermediate particles
may be affected  by the rescattering of outgoing protons. 
An approach similar to Eq.\ref{massx02}, applied 
to elastic rescattering of protons, yields\cite{d136}
\begin{equation}
m{^\prime}_x^2-m_x^2=-2p_bp_f{\sin^2 \phi}
(-{cos ({\theta_b+ \theta_f})}+(2m_N)^{-2}p_bp_f),
\label{rescattering}
\end{equation}
where $p_{b,f}$ are the momenta of protons and  $\phi$ is the
rescattering angle
in the laboratory system.
Thus the contributions from both Fermi motion and from rescattering
(at $\overline{\phi}$$\approx$0.5) in principle 
may be the cause of the observed displacement of the isobar peak.
%
Another option, not investigated yet,
may  relate to the $P_{33}$ nature of the isobar.
The point is that, perhaps, isobar formation 
may be enhanced for slow $S$-wave pions  due to their  
interaction  with  $P$-shell nucleons carrying the desired  angular momentum.

It should be emphasized that unlike the case of   isobar recombination 
process, the effects of Fermi motion and rescattering   may be essential    
for the meson absorption mechanism. Within this approach $t$ is an 
$NN$-pair, and  Eq.\ref{deltamass2} takes the form
\begin{equation}
\langle{m{^\prime}_{x}^2}\rangle-m_{x}^2=
2\mid\varepsilon_{NN}\mid(T_b+T_f)-\langle\hat
{p}_{NN}^2\rangle+\varepsilon_{NN}^2.
\label{deltamass4}
\end{equation}
Assuming that the target nucleons are uncorrelated, we estimate
$\langle\hat{p}_{NN}^2\rangle$$\approx$ $2\langle\hat{p}_{N}^2\rangle$ 
and   $\varepsilon_{NN}$ $\approx$ $2\varepsilon_N$.
Due to  the negative term   -$\langle\hat{p}_{NN}^2\rangle$, 
 Eq.\ref{deltamass4} yields
~~~$(-4.9\pm0.14)$$10^{-2}$GeV$^2$ and $(0.1\pm0.57)$$10^{-2}$GeV$^2$ for 
$P$- and $S$-shells respectively.
Hence, for the  reasons mentioned above, one may expect  that  the 
 major fraction of the measured events\footnote{Unless only $S$-shell nucleons
are involved in the process.} would  
display a  negative $m_x^2$.

However, the data analyzed in the framework of the meson absorption 
hypothesis shows a very surprising behavior.
The $\theta_f$ dependence  of the  yield corresponding to $m_{x}^2$$>$$0$ 
is shown in Fig.\ref{angular_dependence} by the open squares.
It follows the angular dependence of the isobar recombination yield (filled circles).
At the same time, 
the fractional yield, measured as the ratio of   
positive   $m{}_{x}^2$ yield  to the total yield (filled squares),
rises rapidly from 5$\%$  to 92$\%$ (open circles).
The latter value is quite surprising since, contrary to our expectation 
from  Eq.\ref{deltamass4},
most of the events do exhibit positive $m_{x}^2$.

As for the events with negative $m_{x}^2$, we note that in principle
the squared four-momentum of the virtual meson, mediating the 
quasi-free process 
$p+(pn)$$\to$$p_b+p_f+n$ on  $pn$-pair,
has to be negative.  The point is that in such a process the   virtual 
meson  is  emitted by  the projectile proton. 
Therefore the squared mass of intermediate meson  may be evaluated as 
$m_{x}^2$=$(P-P_n)^2$,
where $P$ and $P_n$ are the four-momenta
of the initial proton ($p$) and of the  scattered neutron ($n$) after   
emission of a virtual pion. The resulting value 
is obviously negative since $(P-P_n)^2$=-2$m_p$$T_n$,
where $T_n$ is the kinetic energy of the final neutron in the 
rest frame of the projectile. 
Hence,  
the high relative yield of events with $m_{x}^2$$<$$0$ at 
low $\theta_f$ may be explained
by dominance of the process of proton scattering off 
nucleon pairs, 
which is an essential feature of the model\cite{fs1}. 
The observed sharp build up  of the positive $m_{x}^2$ yield at higher 
$\theta_f$ implies, apparently, that in these events,
prior to ultimate absorption, 
the  meson has its momentum changed via interactions
with other target nucleons. 

\begin{figure}
\begin{center}
\epsfig{file= 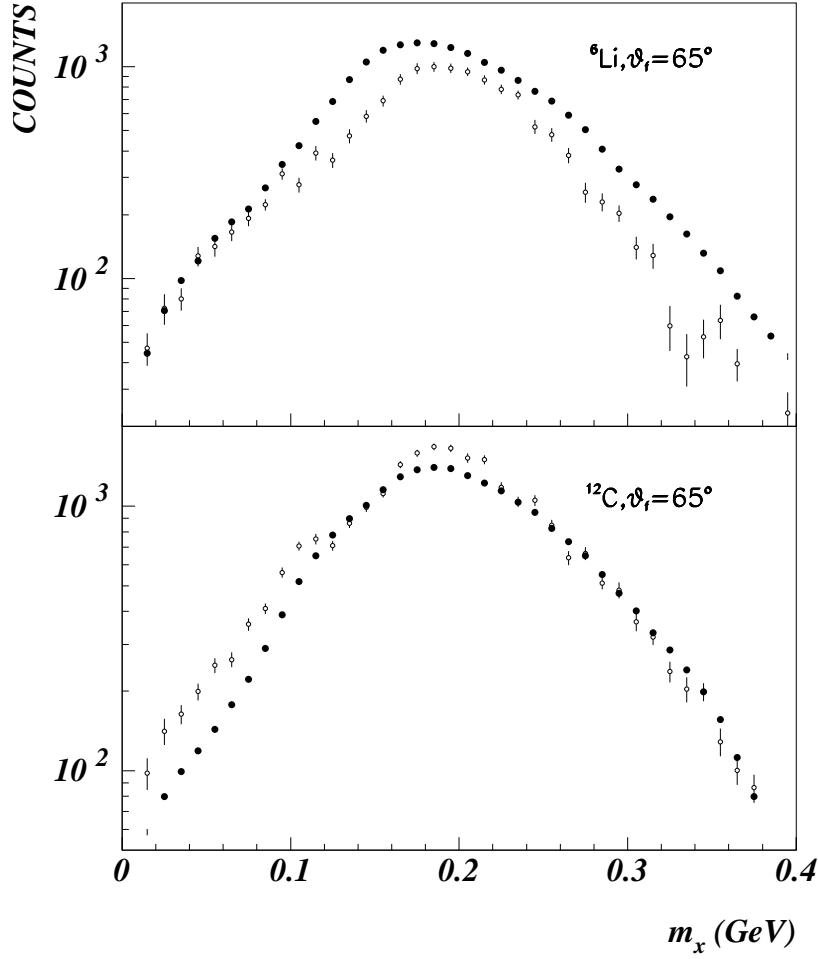,
bbllx=0pt,bblly=128pt,bburx=595pt,bbury=765pt,width=.9\textwidth}
\end{center}
\caption{The mass spectra of intermediate pions at $\theta_f=65^\circ$ 
from   $^{6}Li$ (top panel)  and   $^{12}C$ (bottom panel) targets.
Open circles are the  raw mass spectra evaluated from the momenta 
of $b$- and $f$-protons.
The black dots represent the combinatoric spectra evaluated from 
the pairs of inclusive $b$- and $f$-protons. 
Normalization of the combinatoric spectra is arbitrary.
The statistical errors are less than the symbol size.}
\label{pimass_acc}
\end{figure}

\begin{figure}
\begin{center}
\epsfig{file=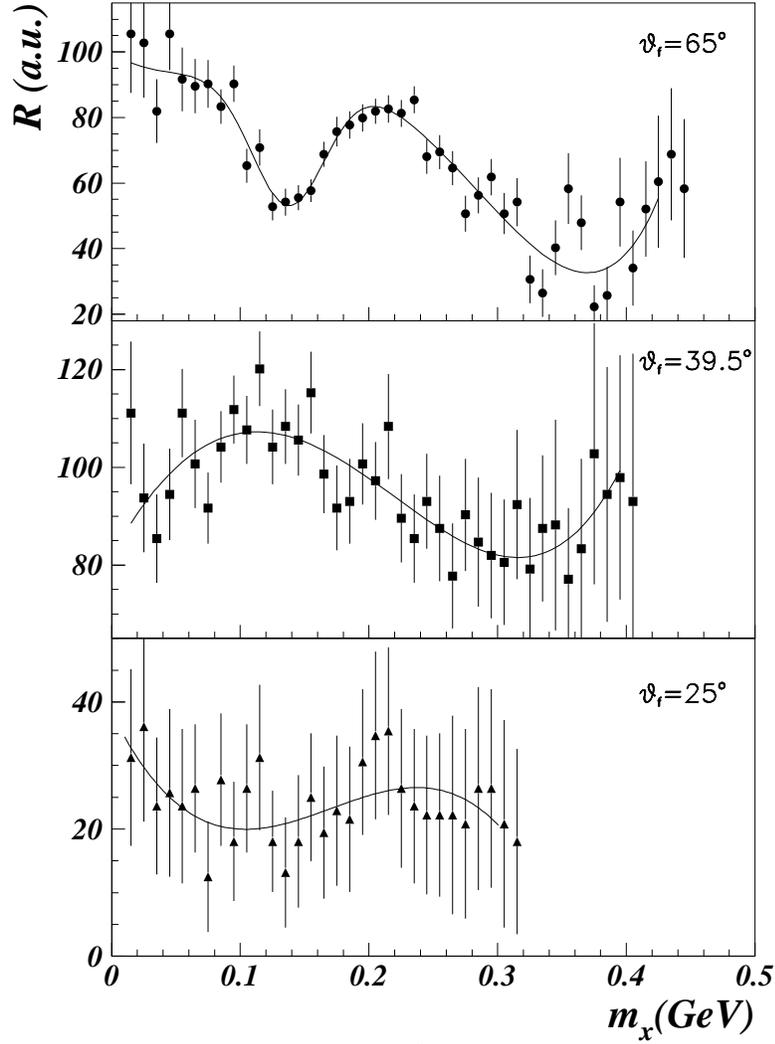,
bbllx=0pt,bblly=128pt,bburx=595pt,bbury=765pt,width=.9\textwidth}
\end{center}
\caption{Correlation function $R$ on $^{6}Li$  
vs the mass of virtual meson ($m_x$),  
calculated within the  meson  absorption  hypothesis at $\theta_f=$~
(1)$65^\circ$(circles),
(2)$39.5^\circ$(squares) and 
(3)$25^\circ$(triangles; the last 15 points are smoothed).
Solid curves  are the   fits  by a Gaussian plus polynomials.}
\label{f_li_pionmass}
\end{figure}

We now consider more closely  the events  with positive $m_{x}^2$. 
For such events  the spectra of reduced mass of the intermediate meson
have been  
investigated within the meson absorption hypothesis.
The samples of  original mass spectra
are shown in  Fig.\ref{pimass_acc}  together with the combinatoric
spectra. 
The resulting reduced spectra  from $^6Li$ at three emission 
angles of $f$-protons are plotted in Fig.\ref{f_li_pionmass}.
$R(m_x)$ at $25^o$ is nearly  constant, while strong
deviations are seen  at $39^o$ and $65^o$.

\begin{figure}
\begin{center}
\epsfig{file=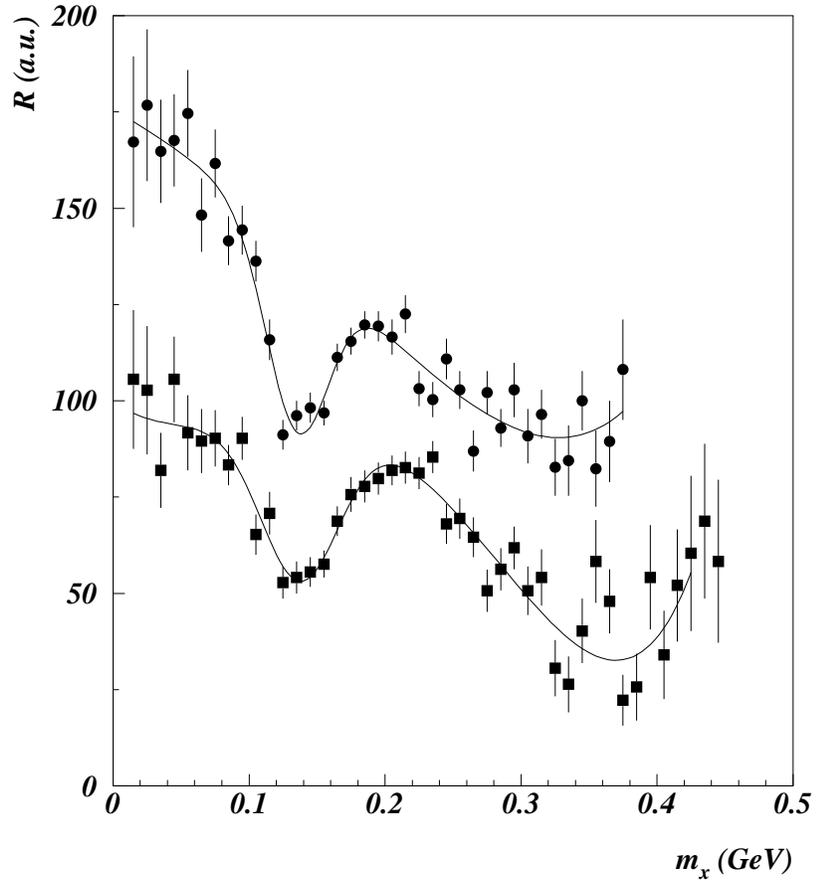,
bbllx=0pt,bblly=128pt,bburx=595pt,bbury=765pt,width=.9\textwidth}
\end{center}
\caption{
The nuclear transparency effect.
The correlation function $R$ at $\theta_f=65^\circ$ 
on targets $^{12}C$(circles) and $^{6}Li$(squares)
vs the mass of the intermediate meson($m_x$).  
The solid curves are the fits by a sum of 
polynomials minus a Gaussian.}
\label{f_lic_pimass}
\end{figure}

The reduced mass spectrum 
at $\theta_f$=$65^o$ is very interesting.
This spectrum exhibits an obvious  dip 
in the vicinity of the pion mass. In order to 
measure the width and location of this dip 
it is natural to fit this  spectrum
 by a sum  of  polynomials minus a Gaussian. 
The fitted centroid of the dip thus obtained, $138\pm2$MeV, agrees 
nicely with  
the  mass of pion, while its standard deviation ($30\pm3$MeV) 
is close to  the mass resolution of the spectrometer, pointing out
that the 
real width of the dip may be  significantly smaller.  
To ensure  this observation, 
we have investigated the  reduced mass spectrum
from $^{12}C$ in comparison  with the $^6Li$ data 
(Fig.\ref{f_lic_pimass}). 
The same narrow 
dip at  $135\pm$2MeV is evident in the $^{12}C$ spectrum.
It is worth mentioning that the  amplitude of the deviation divided by  
$R(0)$  scales as $A^{-1}$ within  $\pm7$$\%$ accuracy, 
and we find it encouraging to see 
a noticeable $\theta_{f}$- and $A$-dependence of the spectrum shape. 
So this  effect is obviously not an instrumental artifact.  

We can suggest the following,
perhaps  naive interpretation of this previously unknown  effect.
In the framework of the  meson absorption hypothesis  
we probe  the mass of the intermediate off-shell meson ($m_{x}$)
in a wide range.
As $m_{x}$ approaches  the mass  of  real pion
from either side, the probability of absorption from 
the intermediate
state  drops. 
It may well be that such behavior
reflects the obvious fact that
%
a real meson has a chance
to escape from the nucleus into the free space, rather than to be 
absorbed by nucleons,
while the virtual meson has not other
options but to be absorbed inside the nucleus.
Hence, a narrow dip at the  mass of pion  builds up, as if  the   
nuclear medium is more "transparent" for real mesons.


It is evident from
the large relative yield of positive $m_{x}^2$ as well as from the dip
location, which
matches perfectly to the  pion mass,
that neither Fermi motion nor rescattering
influence  this  effect.
We emphasize that
it is the term -$\langle\hat{p}_{NN}^2\rangle$ in Eq.\ref{deltamass4} that
may change $m_{x}^2$ to negative values, and,
%
in  contrast with the case of a single  nucleon target,  
%
this term can be even zero for a nucleon pair 
if, for example,  
the couple  populates the same nuclear  shell
with opposite momenta. 
Since the typical momentum transfer in the process of virtual pion absorption
exceeds 
$0.5fm^{-1}$, 
such a pair  could  be  a short-range  $np$-correlation\cite{fs1},
which apparently has been observed recently in  $^{12}C(p,2p+n)$
measurements\cite{e850}.  
%

\section{Conclusion.}
The mechanism of production of cumulative protons 
on light nuclei was studied
implementing the method of correlation functions.
Evidence for the isobar recombination mechanism 
were found  in the  reduced spectra 
of  kinetic energies of secondary protons.
A more profound study was performed by measuring the 
reduced spectra  of the intermediate baryon mass. These spectra  were
found to contain a pronounced peak in the region of
$\Delta(1232)$. 

In order to investigate further the earlier stages of the process, 
the same statistics  were analyzed as the process of     
intermediate  meson absorption by a nucleon pair. 
Two surprising phenomena were revealed within this approach.

The first   is that the fractional yield of  events with 
positive $m^2_x$ builds up rapidly with the angle between the two final
protons.

The second interesting phenomenon is the narrow  dip  in the 
reduced mass spectra of intermediate off-shell mesons, 
positioned at the mass of real pion.
The concept of pion-nuclear transparency 
has been suggested for interpretation of this phenomenon.


In principle, the effect of pion-nuclear transparency may show up  
in collisions of heavy ions at high energy.
If this effect reflects the nuclear shell structure, then the 
condition must hold that the shell structure remains unbroken.
Nuclear shells may be destroyed 
 in any transition of nuclear matter 
into  a new state, such as   the quark-gluon plasma, and, perhaps,
such a transition  could be indicated via this phenomena.
We therefore appeal for  further experimental and theoretical 
 studies  of this interesting effect.
In particular,  it would be very productive  to investigate the
$A$- and $\theta_f$-dependencies of the  transparency dip  
with improved  momentum  resolution.

We are indebted 
to the PNPI synchro-cyclotron crew for the  excellent proton beam.
We  acknowledge L.L.Frankfurt, M.I.Strikman and M.B.Zhalov
for useful discussions and valuable comments on a draft of this paper. 

The preliminary results have been published in \cite{d136,d137,d143}.

\newpage
%
%
%
%
%
%
%
%
%
%
%
%
%

%
%
%
%
%
%
%
%
%
%
%
%


\bibliographystyle{unsrt}
\bibliography{li_sl}

\end{document}